# Research on the Preparation and Luminescent Properties of $Ca_{0.87-x}Sr_xAl_2Si_2O_8$:$0.13Eu^{2+}$ Broad Band Blue Phosphor for White Light LEDs

**Rouziaji Alimu**[a, b, c] **Jiang Yu** [a, b, c] **Wang Maohua** [a, b, c] **Aierken Sidike**[a, b, c,*]

a School of Physics and Electronic Engineering, Xinjiang Normal University, Urumqi, Xinjiang 830054, China.
b Xinjiang Key Laboratory for Luminescence Minerals and Optical Functional Materials, Urumqi, Xinjiang 830054, China.
c Xinjiang Key Laboratory for Mineral Luminescent Material and Microstructure, Urumqi, Xinjiang 830054, China.
*Correspondence to [aierkenjiang@sina.com]

## Abstract

White light emitting diodes (LEDs) have become the most widely used lighting devices due to their high luminous efficiency, environmental friendliness, long lifespan, and energy-saving advantages. Currently, the production of white light LEDs faces challenges such as a lack of cyan light emission, which results in a low color rendering index (Ra), as well as issues related to poor thermal stability of the luminescent materials and difficulties in synthesis. This paper employs a high-temperature solid-state method and cation substitution strategy to prepare a broadband blue phosphor $Ca_{0.87-x}Sr_xAl_2Si_2O_8$: $0.13Eu^{2+}$ (x=0-0.87) with a feldspar structure. By varying the concentration of x, phase transitions are achieved, leading to the formation of two luminescent centers. The luminescence intensity of $Ca_{0.87}Al_2Si_2O_8$:$0.13Eu^{2+}$ is enhanced by 2 times, and the internal quantum efficiency (IQE) increases by 24.8%. The internal quantum efficiency of $Ca_{0.27}Sr_{0.60}Al_2Si_2O_8$:$0.13Eu^{2+}$ reaches as high as 84.1%. The strongest emission peak is observed when the concentration of $Sr^{2+}$ reaches 0.6 mol, with a full width at half maximum of 91 nm. The experimental value of the optical band gap of the matrix is 5.74 eV, indicating that the luminescence intensity of the sample $Ca_{0.27}Sr_{0.60}Al_2Si_2O_8$:$0.13Eu^{2+}$ at 150 °C is 92.2% of that at room temperature. $Ca_{0.27}Sr_{0.60}Al_2Si_2O_8$:$0.13Eu^{2+}$ and commercial red phosphor

(Ca,Sr)AlSiN$_3$:Eu$^{2+}$, as well as commercial green phosphor (Sr,Ba)$_2$SiO$_4$:Eu$^{2+}$, have been successfully packaged into white light LED devices using a 365nm near-ultraviolet chip. The color rendering index (Ra) of this white light LED device is 94.3, with a color coordinate of (0.4094, 0.4167) and a correlated color temperature (CCT) of 3611.4K.

## 1.Introduction

In recent years, phosphor-converted white light-emitting diodes (pc-WLEDs) have gradually become a focus of academic research due to their advantages in efficiency, environmental friendliness, energy saving, and long lifespan. Currently, commercially available pc-WLEDs are mainly produced by combining blue InGaN light-emitting chips with yellow YAG:Ce$^{3+}$ phosphors. However, this method has drawbacks such as a poor color rendering index (CRI<75) and excessively high color temperature due to a severe lack of red light components. The most effective way to improve Ra and CCT is to add red-emitting materials, such as Sr2Si5N8:Eu2+ and (Sr,Ca)AlSiN$_3$:Eu$^{2+}$, to the [InGaN chip + YAG:Ce$^{3+}$] system[1-4]. However, there is still a strong blue emission from the blue InGaN chip, and prolonged exposure to blue light can inhibit the secretion of melatonin in humans, leading to health issues such as the formation of cataracts, disruption of circadian rhythms, and mood disorders[5-6]. It is evident that the aforementioned scheme fails to meet the requirements for high-quality white LED lighting. The use of near-ultraviolet (350-410nm) LED chips combined with red, green, and blue phosphors has great development potential, as it can not only achieve excellent warm white light illumination but also effectively reduce blue light radiation from WLED devices. Although there are a certain number of high-performance blue phosphors available in the market, they are still relatively lacking. Taking BaMgAl$_{10}$O$^{17}$:Eu$^{2+}$ (BAM:Eu$^{2+}$) as an example, this commercial blue phosphor exhibits high quantum efficiency under normal temperature conditions, but its luminous intensity significantly decreases in high-temperature environments. In response to these issues, researchers have proposed various strategies to enhance the luminous performance of BAM:Eu$^{2+}$ phosphors, such as carbon coating and calcination in a vacuum atmosphere[7]. However, it still cannot improve the issue of thermal quenching. To address this, researchers have designed many novel and efficient blue-emitting phosphors. Kim[8] et al. reported a blue-emitting $Na_{3-2x}Sc_2(PO_4)_3:xEu^{2+}$ phosphor, which exhibits zero thermal quenching performance. However, the relatively weak absorption of this phosphor material in the near-ultraviolet range significantly limits its practical application when driven by near-ultraviolet LED chips. Therefore, developing efficient blue phosphors that can be effectively excited by near-ultraviolet LED chips remains a significant challenge. Many rare earth ions can serve as activators and have high luminous efficiency, making rare earth ions commonly used as activators. Eu2+, belonging to the lanthanide series, has been widely studied as an excellent luminescent activator in many matrices. Depending on the crystal environment, the emission resulting from the 4f65d1→4f7 transition of Eu2+ varies from the ultraviolet region to the red light region, making it one of the most ideal activators[9-11]. The model of anorthite ($CaAl_2Si_2O_8$) has a relatively symmetrical crystal structure. In this work, a blue luminescent material doped with $Eu^{2+}$ in $CaAl_2Si_2O_8$ was first synthesized, and the concentration of $Sr^{2+}$ was further increased to replace $Ca^{2+}$ in the matrix. This cation substitution strategy achieved a phase transition, enhancing the luminescence intensity of the sample, while also obtaining a broadband blue phosphor with high quantum yield and thermal stability. A detailed study was conducted on the crystal structure, morphology, photoluminescence, thermal stability, and quantum yield of the resulting phosphor. Additionally, it was combined with near-ultraviolet chips and commercial green and red phosphors to manufacture full-spectrum white light LED devices. This work provides significant reference value for the study of blue luminescent materials.

## 2. Experiment section

### *2.1. Sample preparation*

In this article, $Ca_{0.87-x}Sr_xAl_2Si_2O_8:0.13Eu^{2+}$ phosphors were prepared using a high-temperature solid-state method. The raw materials used include $SrCO_3$ (A.R.), SiO2 (99.99%), $Al_2O_3$ (A.R.), $CaCO_3$ (99.99%), and $Eu_2O_3$ (99.99%). All the raw materials were sourced from Shanghai Aladdin Biochemical Technology Co., Ltd. The amounts of each component were calculated based on stoichiometric molar ratios, weighed using an electronic balance, and then placed in an agate mortar for grinding for 20-25 minutes. The sample was then transferred to an alumina crucible. In a tube furnace under a reducing atmosphere of 1:9 ($H_2:N_2$), the temperature was raised to 1300°C at a heating rate of 5°C/min and maintained for 210 minutes. After cooling to room temperature, the sample was removed and ground in the agate mortar for 20 minutes.

## 2.2. Sample characterization

This experiment used an X-ray diffractometer (XRD, XRD6100 Shimadzu, Japan) to test the phases of the materials. The instrument operated at a current of 30 mA and a voltage of 40 kV, with a scanning rate of 5°/min. The continuous scanning range was selected from 10° to 80°. For samples undergoing structural refinement, the scanning speed was set to 2°/min. A JEOL JSM-7610FPLUS (Japan) scanning electron microscope was used to scan the samples for electron microscope (SEM) images. The composition and elemental distribution of the luminescent materials were analyzed using an X-MaxN50 energy-dispersive X-ray spectrometer (Oxford Instruments, UK). The excitation and emission spectra, lifetime decay curves, and temperature-dependent emission spectra of the samples were analyzed using a FLS920 steady-state/transient fluorescence spectrometer from Edinburgh, UK. The diffuse reflectance characteristics of the samples were analyzed using a UV-2550PC diffuse reflectance spectrometer (Shimadzu, Japan).

## 2.3. LED packaging

In this study, the $Ca_{0.27}Sr_{0.60}Al_2Si_2O_8:0.13Eu^{2+}$ blue phosphor was prepared and mixed with commercial $(Ca,Sr)AlSiN_3:Eu^{2+}$ red phosphor and commercial green phosphor $(Sr,Ba)_2SiO_4:Eu^{2+}$ (produced by Shenzhen Zhanwanglong Technology Co., Ltd.) to obtain white light LED devices. The testing equipment used was the Rainbow Spectrum HP9000. The mixed samples were combined with silicone gel in a ratio of 1:1:1 (phosphor to silicone gel), and the mixture was evenly coated onto a 365nm (1W) LED chip, followed by heating at 110°C for 10 hours. After complete curing, the performance of the LED devices was tested.

## 3. Results and discussion

### *3.1. Physical phase analysis of the samples*

Figures 1(a, b) show the XRD magnified images of $Ca_{0.87-x}Sr_xAl_2Si_2O_8:0.13Eu^{2+}$ (x=0-0.87) at $2\theta=22°\sim35.8°$. It can be seen that when the concentration of x increases from 0 to 0.60, all diffraction peaks match well with the standard card PDF#89-1459 data for $CaAl_2Si_2O_8$. At x=0.87 mol, the diffraction peaks completely match the standard card PDF#70-1862 for $SrAl_2Si_2O_8$, indicating that the phosphor is a pure phase. As the Sr concentration increases, the main diffraction peak at 28.5° gradually decreases and eventually disappears, while a series of peaks gradually appear at 26°-27.5°-33°, indicating that the crystal structure of the sample evolves from $CaAl_2Si_2O_8$ (ISCD: 86317) to $SrAl_2Si_2O_8$ (ISCD: 4354). The above results demonstrate that we have successfully achieved a phase transition from calcium feldspar to strontium feldspar. Considering the principle of charge balance and the similar ionic radii (the difference in radii of the two ions is less than 30%), it is speculated that Eu2+ may occupy the lattice sites of $Sr^{2+}$ and $Ca^{2+}$, resulting in the phase transition phenomenon [12]. Based on the occupancy situation and the principle of charge balance, the following relationship can be established:

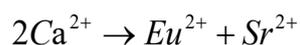

$$2Ca^{2+} \rightarrow Eu^{2+} + Sr^{2+}$$

Figure 1(c) shows the evolution from the $CaAl_2Si_2O_8$ phase to the $SrAl_2Si_2O_8$ phase. $CaAl_2Si_2O_8$ belongs to the triclinic crystal system with a space group of P-12[C]. As can be seen from Figure 1(c), the Ca atom is coordinated with six O atoms to form a $[CaO_6]$ octahedron, while the Al and Si atoms are coordinated with four neighboring O atoms to form $[AlO_4]$ and $[SiO_4]$ tetrahedra. As the concentration of $Sr^{2+}$ ions gradually increases, the originally six-coordinated triclinic structure ($CaAl_2Si_2O_8$) transforms into a monoclinic structure ($SrAl_2Si_2O_8$) with 12/C(15) symmetry. In the crystal structure, the polyhedra $[Ca/SrO_6]$ are connected to $[AlO_4]$ and $[SiO_4]$ through shared vertices and edges. In the lattices of these two samples, there are also two types of O atoms, referred to as O1 and O2. Each $[SiO_4]$ and $[AlO_4]$ tetrahedron shares the top O1 atom, which is cross-located in the framework structure, surrounded by eight $[CaO_6]$ octahedra. The remaining three O2 atoms each share with one [CaO$_6$] octahedron and either an [AlO$_4$] or [SiO$_4$] tetrahedron, thus forming a crystal structure with higher symmetry. As shown in Figure1(c), there is a channel along the x-axis in this structure, and to maintain charge balance within the structure, Ca$^{2+}$ and Sr$^{2+}$ ions are filled into this channel.

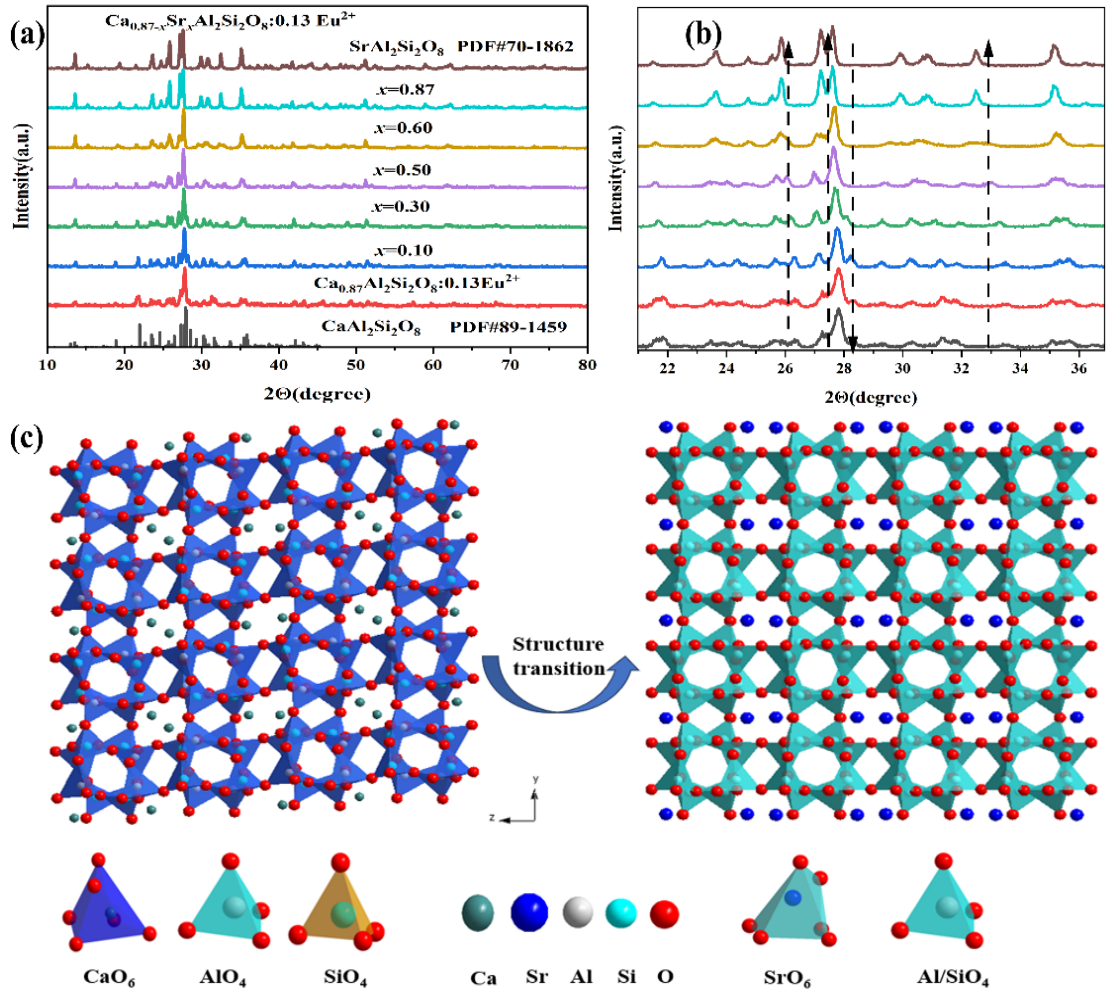

Figure 1 (a) XRD pattern of Ca$_{0.87-x}$Sr$_x$Al$_2$Si$_2$O$_8$:0.13Eu$^{2+}$ (x=0~0.87); (b) Enlarged XRD pattern from 2θ=22° to 36.8°; (c) Evolution of the crystal structure from CaAl$_2$Si$_2$O$_8$ to SrAl$_2$Si$_2$O$_8$.

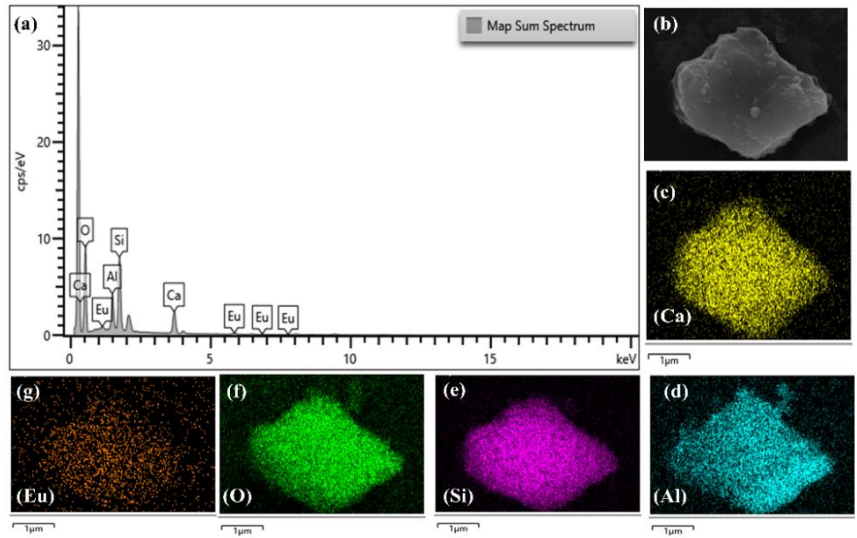

Figure 2 SEM and EDS images of $Ca_{0.87}Al_2Si_2O_8:0.13Eu^{2+}$.

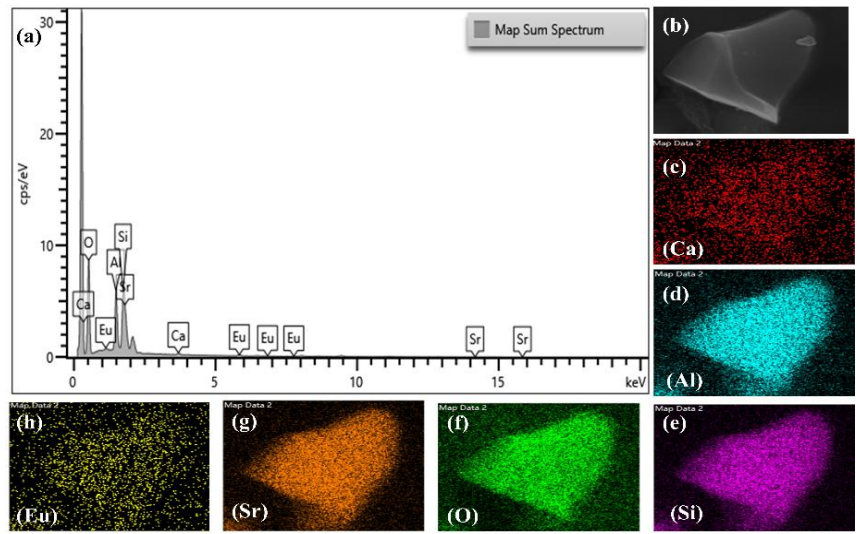

Figure 3 SEM and EDS images of $Ca_{0.27}Sr_{0.60}Al_2Si_2O_8:0.13Eu^{2+}$.

We conducted a simple characterization of the morphology of $Ca_{0.87}Al_2Si_2O_8:0.13Eu^{2+}$ and $Ca_{0.27}Sr_{0.60}Al_2Si_2O_8:0.13Eu^{2+}$, as shown in Figures 2 and 3. From Figure 2, we can see that the surface of $Ca_{0.87}Al_2Si_2O_8:0.13Eu^{2+}$ is not sufficiently smooth. Therefore, we aimed to achieve a sufficiently smooth surface for the phosphor through a cation substitution strategy. Through SEM and EDS analysis, it can be observed from the SEM image in Figure 3(b) that the addition of $Sr^{2+}$ ions has a positive effect on improving the agglomeration phenomenon and the rough surface of the particles in the previous sample. We obtained the desired crystal surface. The morphology of the sample after partial substitution of $Ca^{2+}$ ions with $Sr^{2+}$ ions shows a relatively regular bulk structure, with good dispersion and no agglomeration, and the particle surface is smooth, indicating that the sample has good crystallinity. From Figures 3(c-h), it can be seen that the mapping images of all elements correspond completely with the SEM images, indicating that the elements are uniformly distributed in the sample. Figure 3(a) shows the corresponding EDS spectrum, which clearly indicates the presence of Ca, Sr, Al, Si, O, and Eu elements in the sample, with no other impurity elements detected, demonstrating that this representative sample has excellent purity.

### *3.2. Photoluminescence properties*

As shown in Figure 4(a), the diffuse reflectance spectra of three samples, $CaAl_2Si_2O_8$, $Ca_{0.87}Al_2Si_2O_8:0.13Eu^{2+}$, and $Ca_{0.27}Sr_{0.60}Al_2Si_2O_8:0.13Eu^{2+}$, were tested. $Ca_{0.87}Al_2Si_2O_8:0.13Eu^{2+}$ and $Ca_{0.27}Sr_{0.60}Al_2Si_2O_8:0.13Eu^{2+}$ exhibit a broad absorption band in the range of 250 - 400 nm, indicating that these samples will produce efficient absorption when excited by near-ultraviolet LED chips. Generally, the optical band gap (Eg) of the luminescent material can be roughly calculated from the absorption spectrum of the sample. The optical band gap of the luminescent material can be calculated using the following formula [13]:

$$[F(R\infty)hv]^n = A(hv - E_g)$$

$$F(R\infty) = \frac{(1-R)^2}{2R}$$

Among them, F(R) and A are the absorbance and absorption constant of the sample, respectively, while R and n are the reflectance and transition coefficient, respectively, and hv is the photon energy. The results show that the experimental optical band gaps of $CaAl_2Si_2O_8$, $Ca_{0.87}Al_2Si_2O_8:0.13Eu^{2+}$, and $Ca_{0.27}Sr_{0.60}Al_2Si_2O_8:0.13Eu^{2+}$ are 5.74 eV, 4.6 eV, and 4.64 eV, respectively, as shown in Figures 4(b-d). The relatively wide optical band gap of the samples preliminarily indicates that they have good thermal stability [14].

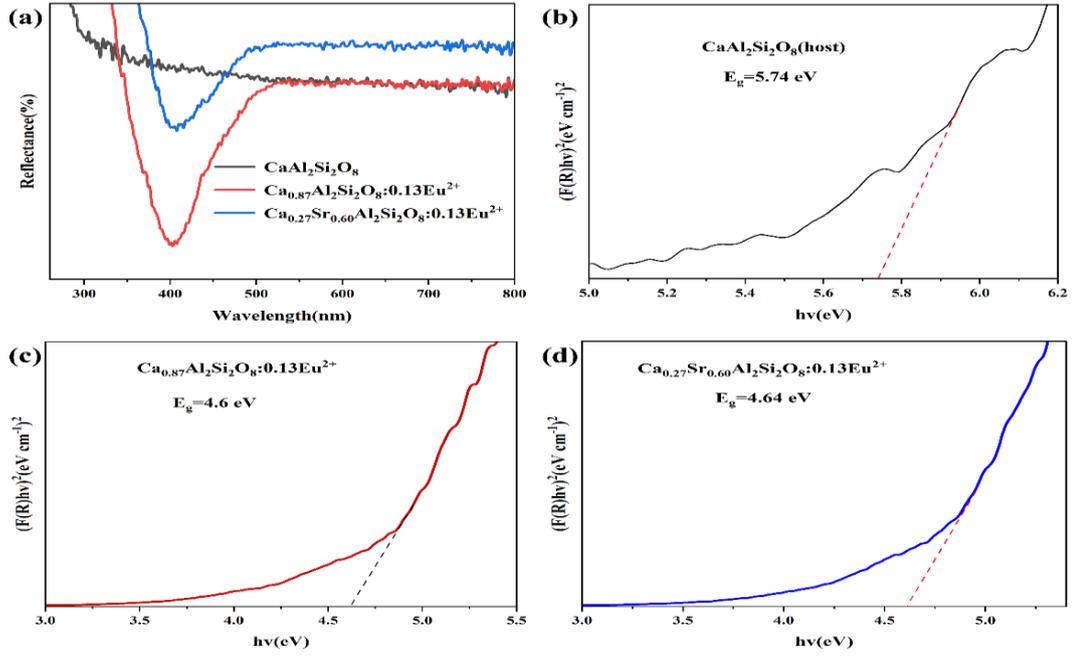

Figure 4 shows the samples $CaAl_2Si_2O_8$, $Ca_{0.87}Al_2Si_2O_8$:$0.13Eu^{2+}$, and $Ca_{0.27}Sr_{0.60}Al_2Si_2O_8$:$0.13Eu^{2+}$. (a) Diffuse reflectance spectra; (b-d) Experimental values of the optical band gap.

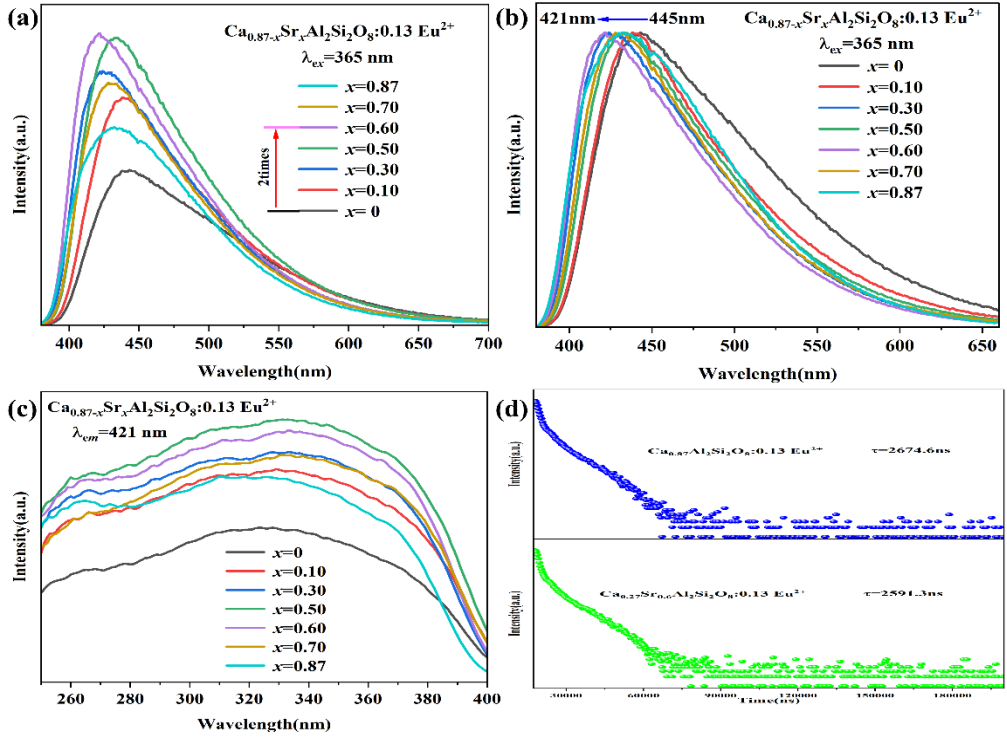

Figure 5 (a) Emission of $Ca_{0.87-x}Sr_xAl_2Si_2O_8$:$0.13Eu^{2+}$ (x=0-0.87); (b) Normalized emission spectrum of $Ca_{0.87-x}Sr_xAl_2Si_2O_8$:$0.13Eu^{2+}$ (x=0-0.87); (c) Excitation spectrum of $Ca_{0.87-x}Sr_xAl_2Si_2O_8$:$0.13Eu^{2+}$ (x=0-0.87) obtained at a detection wavelength of 421 nm; (f) Lifetime decay curves of $Ca_{0.87}Al_2Si_2O_8$:$0.13Eu^{2+}$ and $Ca_{0.27}Sr_{0.60}Al_2Si_2O_8$:$0.13Eu^{2+}$.

We studied the photoluminescent properties of the samples $Ca_{0.87-x}Sr_xAl_2Si_2O_8:0.13Eu^{2+}$ (x=0-0.87) when $Sr^{2+}$ substitutes for $Ca^{2+}$ at different concentrations. As shown in Figure 5, from Figure 5(a), it can be seen that $Ca_{0.87}Al_2Si_2O_8:0.13Eu^{2+}$ exhibits broad-band emission under 365 nm near-ultraviolet excitation, with a full width at half maximum (FWHM) of 116 nm, and the strongest peak of the emission spectrum is located at 443 nm. As the concentration of Sr2+ increases, the emission intensity of the phosphor first increases and then decreases. When the concentration of Sr2+ reaches 0.6 mol, a broad-band emission at 421 nm with a FWHM of 91 nm is obtained. At this point, the luminescence intensity of the sample is the strongest, with the maximum luminescence intensity under 365 nm excitation being twice that of the original. However, when the concentration of $Sr^{2+}$ is between 0.70 and 0.87 mol, the luminescence intensity decreases instead. It can be concluded that a concentration quenching ph enomenon occurs with the increase in $Sr^{2+}$ concentration, which is also verified during the lifetime testing process. The optimal substitution concentration of $Sr^{2+}$ in the matrix is 0.6 mol. The main reason for the enhancement of luminescence intensity may be that after the addition of $Sr^{2+}$, $Eu^{2+}$ occupies not only the $Ca^{2+}$ sites but also the $Sr^{2+}$ sites. As the concentration of $Sr^{2+}$ increases, the amount of $Eu^{2+}$ occupying the Sr2+ lattice sites also increases. The relative distance between each $Eu^{2+}$ becomes smaller than the critical distance, enhancing the interactions between them, which ultimately leads to energy transfer. This is the reason for the concentration quenching phenomenon occurring in the matrix of this study.

The normalized emission spectrum shows that as the concentration of $Sr^{2+}$ increases, the emission spectrum of the sample exhibits a blue shift. We summarize the reason for the blue shift as a change in the crystal field environment around $Eu^{2+}$, which leads to the splitting of the 5d energy level of $Eu^{2+}$. The degree of crystal field splitting can be estimated using the following formula [15]:

$$D_q = \frac{1}{6} Ze^2 \frac{r^4}{R^5}$$

Among them, Z and Dq represent the degree of splitting of the 5d energy level and the cation valence state or charge, respectively. e and r are the charge of the electron and the radius of the d wave function, respectively, while R is the bond length between the central atom and the coordinating atom. Generally, Z, e, and r are constants, so the crystal field splitting degree mainly depends on the bond length R. From the formula, it can be seen that Dq is inversely proportional to R, meaning that as R increases, Dq decreases, resulting in a shorter emission wavelength and leading to high-energy emission. In this study, when $Sr^{2+}$ partially replaces $Ca^{2+}$, the bond length of the central atom increases, which weakens the crystal field strength around $Eu^{2+}$, reduces the splitting degree Dq, and leads to high-energy emission, causing the wavelength emitted by $Eu^{2+}$ to shorten. This is also the reason for the blue shift phenomenon.

Figure 5 (c) shows the excitation spectrum of a series of samples doped with $Sr^{2+}$ ions at a detection wavelength of 421 nm. It can be observed that the excitation intensity at 365 nm increased by 1.6 times after $Sr^{2+}$ ions partially replaced $Ca^{2+}$ ions. The sample exhibits a very broad excitation band in the 250-400 nm range, with two excitation peaks at 264 nm and 309 nm, which correspond to transitions between different split energy levels of the $Eu^{2+}$ 4f to 5d energy levels. The results indicate that this phosphor can be effectively excited by near-ultraviolet and ultraviolet LED chips.

When adopting a cation substitution strategy, multiple luminescent centers may sometimes be generated. The main reason for this is related to the doped ions and the ions being substituted. If the doped ions partially replace the ions that the activator is supposed to replace, it may lead to the formation of multiple luminescent centers. However, if the substitution occurs in the surrounding ionic sites, it generally does not produce multiple luminescent centers. In this experiment, since $Sr^{2+}$ partially replaced the $Ca^{2+}$ sites, and $Eu^{2+}$ originally occupied the $Ca^{2+}$ sites, after $Sr^{2+}$ partially replaced the $Ca^{2+}$ sites, $Eu^{2+}$ could occupy both the $Ca^{2+}$ and $Sr^{2+}$ sites. As a result, two luminescent centers appeared. Moreover, the phenomenon of generating two luminescent centers has a positive effect on the luminescent performance of the phosphor. Further testing and verification of the two luminescent centers in the phosphor $Ca_{0.27}Sr_{0.60}Al_2Si_2O_8:0.13Eu^{2+}$ will be conducted. Figure 5 (d) shows the lifetime decay curves of $Ca_{0.87}Al_2Si_2O_8:0.13Eu^{2+}$ and $Ca_{0.27}Sr_{0.60}Al_2Si_2O_8:0.13Eu^{2+}$,

which can be well-fitted using a double exponential function. Their lifetime values can be calculated using the following formula [16]:

$$\tau^* = \frac{A_1\tau_1^2 + A_2\tau_2^2}{A_1\tau_1 + A_2\tau_2}$$

In the formula, the short lifetime and long lifetime are represented by the symbols, while A1 and A2 are the corresponding constant factors for the short and long lifetimes, respectively. By combining the formula with experimental data, the following results can be obtained: with the concentration of $Eu^{2+}$ ions remaining constant, as the concentration of $Sr^{2+}$ ions increases, there is a significant difference in the corresponding lifetime values. The fluorescence lifetimes of $Ca_{0.87}Al_2Si_2O_8:0.13Eu^{2+}$ and $Ca_{0.27}Sr_{0.60}Al_2Si_2O_8:0.13Eu^{2+}$ are 2674.6 ns and 2591.3 ns, respectively. Based on these two different lifetime values, it is initially believed that this phosphor may have multiple luminescent centers.

Figure 6(a) shows the representative sample $Ca_{0.27}Sr_{0.60}Al_2Si_2O_8:0.13Eu^{2+}$ with the strongest luminescence intensity. It was subjected to Gaussian fitting under 365 nm excitation, as shown in Figure 6(b). Through Gaussian fitting, it can be observed that this representative sample can fit two symmetric Gaussian peaks with energies of 2.7 eV and 3.01 eV, indicating the presence of two luminescent centers, with the peak at 2.7 eV being clearly dominant. According to the principle of similar ionic radii, $Eu^{2+}$ ions occupy the lattice sites of $Ca^{2+}$ and $Sr^{2+}$ ions. To further determine the occupancy of $Eu^{2+}$ in the crystal, the energy (E) of the emission spectrum can be calculated using the following Van Uitert empirical formula [17].

$$E(cm^{-1}) = Q \times \left[1 - \left(\frac{v}{4}\right)^{\frac{1}{v}} 10^{-\frac{nrEa}{80}}\right]$$

In this context, Q and v represent the energy of the lowest point of the low-energy level of $Eu^{2+}$ and the valence state of the ion (2+), respectively. n and r are the coordination number of the substituted cation and the ionic radius of the substituted oxygen ion, respectively. Ea is the electron affinity of the anion in the sample (which is generally constant within the same matrix). r(Ca) = 0.100 (CN=6); r(Sr) = 0.118

(CN=6). Ea and v are constants in this matrix, so the value of nr determines the value of energy (E). The larger the nr value, the greater the E value, resulting in a shorter wavelength; the smaller the nr value, the smaller the E value, resulting in a longer wavelength [18]. Based on the values of n and r provided above, it can be concluded that the nr value of $Sr^{2+}$ is greater than that of $Ca^{2+}$, indicating that E(Sr) > E(Ca). The Gaussian peak at the long wavelength of 400 nm corresponds to Eu(1) occupying the Ca2+ lattice site, with an energy value of 2.7 eV; the Gaussian peak at the short wavelength of 465 nm corresponds to Eu(2) occupying the $Sr^{2+}$ lattice site, with an energy value of 3.01 eV.

The occupancy of Eu2+ in the matrix can also be determined by testing the lifetime of each Gaussian peak. Figures 6(c, d) show the lifetime decay curves of the two Gaussian peaks at 460 nm and 435 nm for the sample $Ca_{0.27}Sr_{0.60}Al_2Si_2O_8:0.13Eu^{2+}$ under 365 nm excitation. The lifetime of the phosphor fits well with a double exponential function, with the lifetime values of the two Gaussian peaks being 2498.9 ns and 1769.7 ns, respectively. The inconsistency in the lifetimes of the two Gaussian peaks suggests that $Eu^{2+}$ occupies two different lattice sites in $Ca_{0.27}Sr_{0.60}Al_2Si_2O_8:0.13Eu^{2+}$.

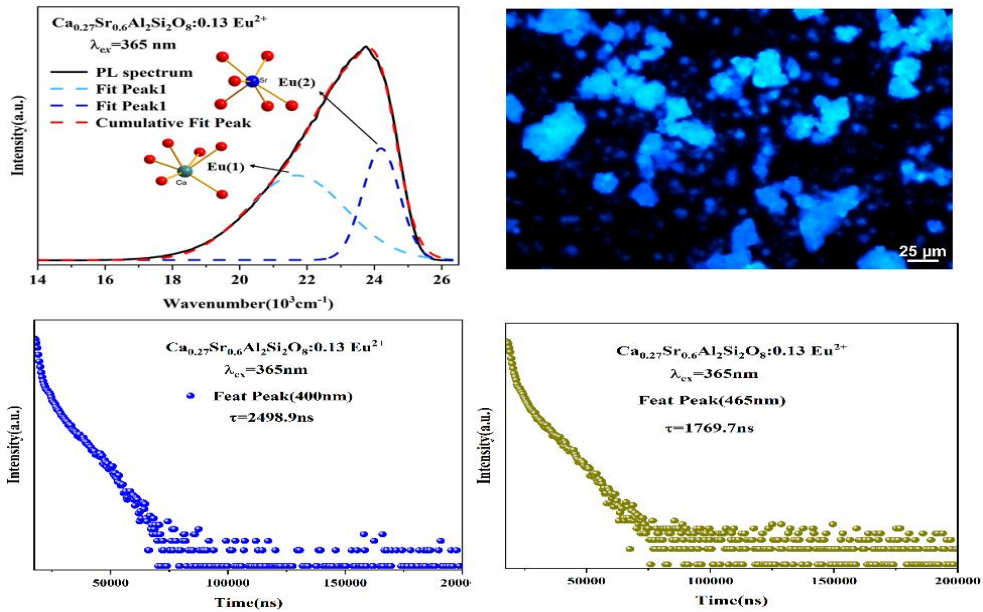

Figure 6 (a) Gaussian fitting of the emission spectrum of $Ca_{0.27}Sr_{0.60}Al_2Si_2O_8:0.13Eu^{2+}$; (b) fluorescence microscopy image; (c, d) lifetime graphs corresponding to Gaussian peaks FitPeak1 (400nm) and FitPeak2 (450nm) of $Ca_{0.27}Sr_{0.60}Al_2Si_2O_8:0.13Eu^{2+}$.

Quantum efficiency is an important indicator for evaluating the luminescent performance of phosphors [19]. Therefore, the internal quantum efficiency of $Ca_{0.87}Al_2Si_2O_8:0.13Eu^{2+}$ blue phosphor was tested. Figures 7 (a,b) show the internal quantum efficiency test results for the samples $Ca0.87Al2Si2O8:0.13Eu2+$ and $Ca_{0.27}Sr_{0.60}Al_2Si_2O_8:0.13Eu^{2+}$, with the inset displaying an enlarged view of their emission spectra. Under 365 nm excitation, the internal quantum efficiency of the phosphor without cation substitution strategy was 59.3%; when $Sr^{2+}$ partially replaced $Ca^{2+}$, the internal quantum efficiency reached 84.1%. Table 1 presents relevant data on the excitation and emission wavelengths and internal quantum efficiency of blue phosphors in recent years. From the comparative results, $Ca_{0.27}Sr_{0.60}Al_2Si_2O_8:0.13Eu^{2+}$ exhibits a relatively excellent internal quantum efficiency, making it a potential material to supplement the blue light component of white light LEDs.

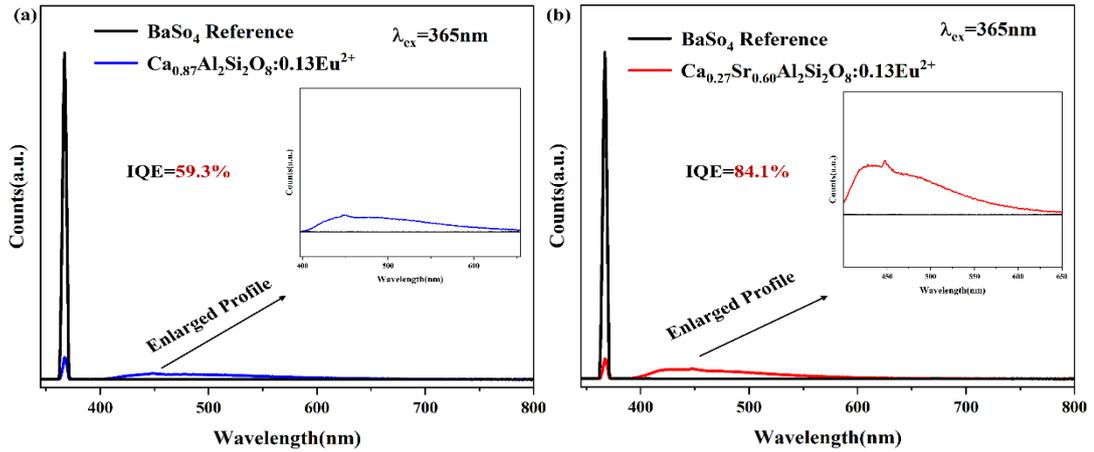

Figure 7 (a, b) Quantum efficiency test graphs for $Ca_{0.87}Al_2Si_2O_8:0.13Eu^{2+}$ and $Ca_{0.27}Sr_{0.60}Al_2Si_2O_8:0.13Eu^{2+}$

Table 1: The excitation/emission wavelengths and internal quantum efficiency of blue phosphors reported in recent years.

| Types of phosphors | Ex(nm) | Em(nm) | IQE | References |
|---|---|---|---|---|
| $Ca_6BaP_4O_{17}: Ce^{3+},Si^+$ | 400 | 480 | 70% | [20] |
| $BaY_2Si_6O_{24}: Ce^{3+}$ | 394 | 480 | 60% | [21] |
| $Sr_5(PO_4)_3Cl: Eu^{2+}$ | 395 | 444 | 81% | [22] |
| $BAON_{1.0}: Eu^{2+}$ | 400 | 468 | 80% | [23] |
| $Ca_{0.27}Sr_{0.60}Al_2Si_2O_8:0.13Eu^{2+}$ | 365 | 421 | 84% | This work |

### 3.3. Thermal stability analysis

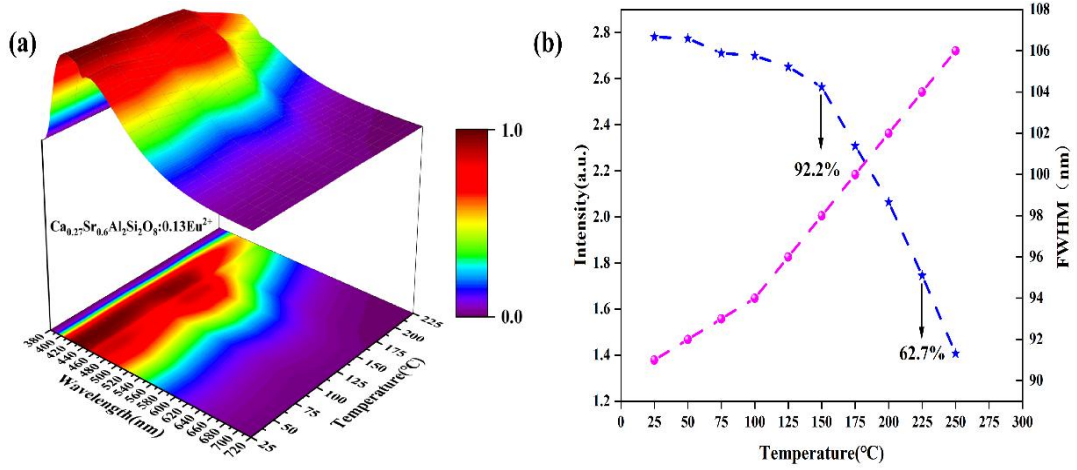

Figure 8 shows the emission spectra of the $Ca_{0.27}Sr_{0.60}Al_2Si_2O_8:0.13Eu^{2+}$ sample (a) at different temperatures; (b) the luminescence intensity at different temperatures.

White light LED devices typically operate in high-temperature environments, so high thermal stability is particularly important among the many parameters of phosphors used in white light LEDs. In this article, the representative sample $Ca_{0.27}Sr_{0.60}Al_2Si_2O_8:0.13Eu^{2+}$ was selected, and the relationship between its temperature and photoluminescent properties was studied in detail. The research found that as the temperature increases, the luminescence intensity of the sample gradually decreases. Figure 8 shows the temperature-dependent emission spectra and point-line graphs of $Ca_{0.27}Sr_{0.60}Al_2Si_2O_8:0.13Eu^{2+}$ from 25°C to 250°C. From Figures 8(a, b), it can be concluded that when the temperature of the representative sample $Ca_{0.27}Sr_{0.60}Al_2Si_2O_8:0.13Eu^{2+}$ reaches 150°C, the luminescence intensity is 92.2% of that at room temperature, and the full width at half maximum of the emission spectrum broadens from 91 nm to 106 nm. The enhancement of electron-phonon interactions at higher temperatures causes the broadening of the emission spectrum. The relationship between the full width at half maximum of the emission spectrum and temperature can be expressed by the following formula [24]:

$$FWHM = W_0 \sqrt{\coth\left(\frac{hv}{2kt}\right)}$$

$$W_0 = \sqrt{8\ln 2}\,(hw)\sqrt{S}$$

In this context, FWHM refers to the full width at half maximum, W0 is the half width of the emission peak at 0K, hw is the energy of lattice vibrations, k is the Boltzmann constant, t is the temperature, and S is the Huang-Rhys-Pekar constant. Generally, the better the stability of the crystal structure, the smaller the lattice vibrations; conversely, the worse the stability of the crystal structure, the larger the lattice vibrations. Temperature has a certain impact on the stability of the crystal structure of luminescent materials[25]. At high temperatures, the stability of the crystal structure deteriorates, the structure becomes loose, lattice vibrations are enhanced, and ultimately this leads to broadening of the emission spectrum. The results show that we further demonstrate that the $Ca_{0.27}Sr_{0.60}Al_2Si_2O_8:0.13Eu^{2+}$ phosphor has good thermal stability.

### 3.4. Packaging and testing

In order to further investigate the potential applications of $Ca_{0.27}Sr_{0.60}Al_2Si_2O_8:0.13Eu^{2+}$ phosphor, white light LED devices were fabricated using near-ultraviolet LED chips (365nm), commercial broadband green phosphor $(Sr,Ba)_2SiO_4:Eu^{2+}$, commercial broadband red phosphor $(Sr,Ca)AlSiN_3:Eu^{2+}$, and the broadband blue phosphor $Ca_{0.27}Sr_{0.60}Al_2Si_2O_8:0.13Eu^{2+}$ prepared in this study. The emission spectrum of the LED devices was tested under a driving current of 30mA, as shown in Figure 9. The device had a correlated color temperature (CCT) of 3611.4K, a color rendering index (Ra) of 94.3, and color coordinates of (0.4094, 0.4167), emitting bright white light.

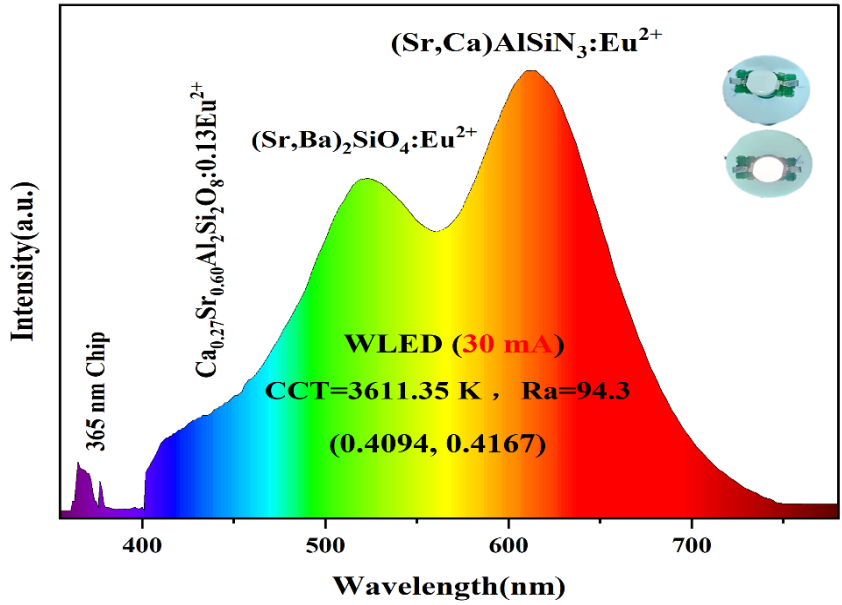

Figure 9 shows the emission spectrum of a white LED device at a current of 20mA; the inset is a picture of the LED device.

## 4. Conclusion

Blue phosphors with a feldspar structure, $Ca_{0.87-x}Sr_xAl_2Si_2O_8:0.13Eu^{2+}$ (x=0-0.87), were prepared using a high-temperature solid-state method and cation substitution strategy. By varying the concentration of x, a phase transition was achieved, resulting in the formation of two luminescent centers. The luminescence intensity of $Ca_{0.87}Al_2Si_2O_8:0.13Eu^{2+}$ was enhanced by 2 times, and the internal quantum efficiency (IQE) increased by 24.8%. The internal quantum efficiency of Ca0.27Sr0.60Al2Si2O8:0.13Eu2+ reached as high as 84.1%. The strongest emission peak was observed when the concentration of Sr2+ reached 0.6 mol, with a full width at half maximum of 91 nm. The experimental value of the optical band gap of the matrix was 5.74 eV, indicating that the luminescence intensity of the sample $Ca_{0.27}Sr_{0.60}Al_2Si_2O_8:0.13Eu^{2+}$ at 150 ° C was 92.2% of that at room temperature. $Ca_{0.27}Sr_{0.60}Al_2Si_2O_8:0.13Eu^{2+}$ was successfully combined with commercial red phosphor $(Ca,Sr)AlSiN_3:Eu^{2+}$ and commercial green phosphor $(Sr,Ba)_2SiO_4:Eu^{2+}$ to package white light LED devices using a 365 nm near-ultraviolet chip. The Ra of the white light LED device was 94.3, with a color coordinate of (0.4094, 0.4167) and a correlated color temperature (CCT) of 3611.4K.